\documentclass[preprint]{emulateapj} 

\usepackage{epsfig,graphicx,latexsym,amsmath,amssymb}
\usepackage{natbib}
\usepackage{hyperref}
\usepackage{breakurl}
\usepackage{mathrsfs}
\usepackage{lastpage}
\usepackage{color}

\renewcommand\vec{\mathbf}
\def\na{{\it New Astron. }}

\bibpunct[,]{(}{)}{;}{a}{}{,}

\begin{document}

\title{Stability of Gas Clouds in Galactic Nuclei: An Extended Virial Theorem}

\author{
Xian Chen\altaffilmark{1}\thanks{},
Pau Amaro-Seoane\altaffilmark{2}\thanks{} \&
Jorge Cuadra\altaffilmark{1}\thanks{}
}

\altaffiltext{1}{Instituto de Astrof\'{i}sica, Facultad de F\'{i}sica, Pontificia Universidad Cat\'{o}lica de Chile, 782-0436 Santiago, Chile; E-mail: xchen@astro.puc.cl}

\altaffiltext{2}{Max Planck Institut f\"ur Gravitationsphysik
(Albert-Einstein-Institut), D-14476 Potsdam, Germany; E-mail: Pau.Amaro-Seoane@aei.mpg.de}


\begin{abstract}

Cold gas entering the central $1$ to $10^2$ pc of a galaxy fragments and
condenses into clouds. The stability of the clouds determines whether they will
be turned into stars or can be delivered to the central supermassive black hole
(SMBH) to turn on an active galactic nucleus (AGN).  The conventional criteria
to assess the stability of these clouds, such as the Jeans criterion and Roche
(or tidal) limit, are insufficient here, because they assume the dominance of
self-gravity in binding a cloud, and neglect external agents, such as pressure
and tidal forces, which are common in galactic nuclei.  We formulate a new
scheme for judging this stability.  We first revisit the conventional Virial
theorem, taking into account an external pressure, to identify the correct
range of masses that lead to stable clouds.  We then extend the theorem to
include an external tidal field, crucial for the stability in the region of
interest -- in dense star clusters, around SMBHs.  We apply our {\em extended
Virial theorem} to find the correct solutions to practical problems that until
now were controversial, namely, the stability of the gas clumps in AGN tori,
the circum-nuclear disk in the Galactic Center, and the central molecular zone
of the Milky Way.  The masses we derive for these structures are orders of
magnitude smaller than the commonly-used Virial masses (equivalent to the Jeans
mass). Moreover, we prove that these clumps are stable, contrary to what one
would naively deduce from the Roche (tidal) limit.

\end{abstract}

\keywords{hydrodynamics -- instabilities -- ISM: kinematics and dynamics --
galaxies: active -- galaxies: nuclei}

\maketitle

\section{Introduction}

An essential ingredient in galaxy formation and evolution is the impact of the
growth of supermassive black holes (SMBHs) on the galactic environment
\citep{kormendy13}.  The agent is gas -- a small amount of cool gas with low
angular momentum fuels the central SMBH, turning on an active galactic nucleus
(AGN), which by means of radiation and outflow heats the gas content on larger,
galactic scales, hence regulating the star formation rate in the entire galaxy
and possibly cutting off the gas supply to the AGN
\citep{mcnamara07,fabian12}.

A bottleneck occurs when cold gas is funneled to the nuclear region of a galaxy
and forms a disk (because of angular-momentum conservation), at a distance of a
few hundreds to several parsec (pc) from the SMBH. At this stage, the gas
density rises to the critical point of triggering the Jeans instability, and
hence the gaseous disk fragments and becomes clumpy
\citep{shlosman90,goodman03}.  From this point on, the inflow of gas is carried
on mainly by clouds. But it is known that a cloud produced by the Jeans
instability is self-gravitating, and will collapse on the free-fall timescale.
The consequence is that within a short time most clouds would turn into stars,
so essentially, and contradictory to many observations, little gas is left to
fuel the central AGN \citep{goodman04,tan05,collin08}.

To resolve this contradiction, several ideas have been proposed.  In
particular, two possibilities to deliver gas clouds to the AGN in spite of them
being Jeans unstable have been considered. (i) A cloud can be stabilized by a
strong magnetic field or fast rotation \citep{rees87,vollmer01b}. (ii) The
inflow of gas could be accelerated, as a result of an asymmetric gravitational
potential \citep{shlosman90,maciejewski02} or cloud-cloud collisions
\citep{krolik88,kumar99,vollmer04}, so that gas clouds can be delivered to the
vicinity of SMBHs before turning into stars.

An alternative viewpoint is that the gas clouds, after the initial
fragmentation, have gone through a series of physical changes, so that the
eventual physical properties no longer satisfy the Jeans criterion.  Two
physical processes could cause such changes: (i) Big clouds
could break into smaller ones due to rapid cooling \citep{burkert00}.  Each
descendant cloud can be stable because of the reduced self-gravity. A galactic
nucleus is an ideal place of producing such clouds, because the high gas
density there enhances the cooling rate.  (ii) Collisions between gas clouds
are frequent in galactic nuclei, since the number density of cloud is high and
the orbital periods are short \citep{krolik88}.  Such collisions raise the
internal pressure of a cloud, supporting it against gravitational collapse.

It is possible today to test the stability of gas clouds with observational
data.  (i) Clumps have been seen in the central hundreds of pc of many
galaxies, quiescent and active \citep{hunt04,mazzalay13,davies14}. (ii) For
AGNs, it has been widely accepted that a clumpy, dusty torus with a size of
several pc must exist surrounding the SMBH to cause the type I/type II
dichotomy \citep{krolik88,antonucci93}. Indeed, direct evidence of such clumps
has been found by modeling the X-ray variability of AGNs, so that we have a
way of measuring some of the clouds' physical properties  \citep[][and references
therein]{risaliti02,markowitz14}.  (iii) In the Milky Way, about $200$
molecular clumps have been found \citep{miyazaki00,oka01} within a distance of
$200$ pc from the Galactic Center (GC), in the region known as the ``central
molecular zone'' \citep[CMZ, ][]{mezger96}.  Further inside, at a distance of
$1$ to $3$ pc from Sgr A* (the SMBH, $8$ kpc from Earth), there is the
circum-nuclear disk (CND), which has been resolved into about $30$ molecular
clumps whose orbits are mostly circular
\citep{genzel85,jackson93,marr93,marshall95}.

To analyze the stability of the observed clumps, a criterion is required.
However, the conventional Jeans criterion and Roche limit become insufficient,
because they regard self-gravity as the only force binding a gas cloud,  which
is not true in galactic nuclei. Here gas clouds are likely subject to a
compression by a high external pressure. This pressure exists because the
interstellar medium (ISM) in the nuclear region of a galaxy usually has a high
density and a large turbulent velocity.  Moreover, there is likely a strong
tidal force acting onto the clouds, because of the ubiquity of SMBHs and dense
nuclear star clusters \citep[see][ for a review]{ama12}. Here we show that
these agents are crucial in determining whether a cloud is stable, as well as
in the correct derivation of its mass.

The paper is organized as follows.  In Section~\ref{sec:theorem}, we revisit
the Virial theorem so as to link the physical properties of a gas cloud to its
environmental conditions, including the external pressure (Section~\ref{sec:P})
and the background tidal field (Section~\ref{sec:tidal}).  After taking these
external factors into account, we formulate in Section~\ref{sec:eVirial} our
``{\em extended Virial theorem}'', which gives only one stable solution for the
mass, and we discuss its implications.  In the light of this stable solution,
we proceed in Section~\ref{sec:app} to interpret the observational data of gas
clumps, including those in AGN tori (Section~\ref{sec:tori}), the CND
(Section~\ref{sec:CND}), and the CMZ (Section~\ref{sec:CMZ}) of the Milky Way.
Finally in Section~\ref{sec:dis}, we justify the assumptions that we adopted in
the work, and we discuss the implications of the results.

\section{Virial Theorem}\label{sec:theorem}

\subsection{External Pressure}\label{sec:P}

A strong external pressure affects the motion of the fluid at the surface of a
gas cloud, and changes the condition of stability of the entire cloud. Such a
high pressure is found in galactic nuclei because of a concentration of hot
ionized- and turbulent cold gas. The Virial theorem \citep{clausius1870} accounting for an external
pressure has been derived in the past \citep{spitzer78,shu92} and has been
applied to the molecular clouds in the CMZ of the Milky Way
\citep[e.g.][]{miyazaki00,oka01}.  In this subsection we present the derivation
in detail so as to highlight an important list of conceptual points -- which are
often overlooked in the literature -- that will be crucial for the main ideas
of this work.

Without magnetic fields and tidal forces affecting it, the equation of motion
of a fluid element can be written as:

\begin{equation}
\rho\ddot{\vec{r}}=-\vec{\nabla}P-\rho\vec{\nabla}\Phi_c,\label{eq:eom}
\end{equation}

\noindent
where $\rho$ is the fluid density, $\vec{r}$ is its position vector, the dots
denote time derivatives, $P$ is the internal pressure including both thermal
and turbulent motion, and $\Phi_c$ is the gravitational potential
(self-gravity).  If we take dot product with $\vec{r}$ on both sides of the
last equation and integrate it over the entire volume of the cloud, the
left-hand-side (LHS) of Equation~(\ref{eq:eom}) becomes

\begin{align}
\int_V\rho\left(\vec{r}\cdot\ddot{\vec{r}}\right)dV&
=\frac{1}{2}\frac{D^2}{Dt^2}\int_V\rho r^2dV-\int_V v^2\rho dV.\label{eq:intlhs}
\end{align}

\noindent
If the internal structure of the cloud is steady, the first term on the
right-hand-side (RHS) of Equation~(\ref{eq:intlhs}) vanishes because
$\rho\,r^2$ is constant. If the cloud is also not rotating, the second term
vanishes, because $v$, the macroscopic velocity of a fluid element, is zero.

Since we are integrating Equation~(\ref{eq:eom}), when the integral of the LHS
is zero, the integrated RHS must also be zero:

\begin{align}
0&=-\int_V\left(\vec{r}\cdot\vec{\nabla} P+\rho \vec{r}\cdot\vec{\nabla} \Phi_c\right)dV\label{eq:intRHS}\\
&=\int_V\left[P\vec{\nabla}\cdot \vec{r}-\vec{\nabla}\cdot(\vec{r}P)\right]dV
-\int_V(\vec{r}\cdot\vec{\nabla} \Phi_c) dm\\
&=\int_V3PdV-\int_SP_S\vec{r}\cdot d\vec{S}+\int_V\vec{r}\cdot\vec{a}_{\rm gr}dm,\label{eq:virial}
\end{align}

\noindent
where $P_S$ denotes the pressure at the surface of the cloud, $dm=\rho\,dV$ is the
mass of the fluid element, $d\vec{S}$ is the
differential area vector orthogonal to the surface of the cloud, and we have
introduced for convenience $\vec{a}_{\rm gr}\equiv-\vec{\nabla}\Phi_c$, the
self-gravity of the cloud.  We now assume spherical symmetry and homogeneity
for the cloud to simplify the integration. We note however that this assumption does not
have an impact in the general conclusions that we will draw. Homogeneity leads to a constant
density $\rho$ and a constant one-dimensional velocity dispersion $\sigma$
inside the cloud, so $P=\rho\sigma^2$ is also constant. In addition with the assumption of
sphericity, Equation~(\ref{eq:virial}) reduces to

\begin{align}
4\pi R_c^3P_S=3M\sigma^2-aGM^2/R_c,\label{eq:virial1}
\end{align}

\noindent
the {\rm conventional} Virial theorem for a steady, non-rotating cloud embedded in a
pressurized medium. Here $R_c$ is the radius of the cloud, $M$ is the cloud
mass, $G$ is the gravitational constant, and $a$ is a geometrical factor of
order unity, which is $3/5$ assuming sphericity.

\begin{figure}
\includegraphics[width=84mm]{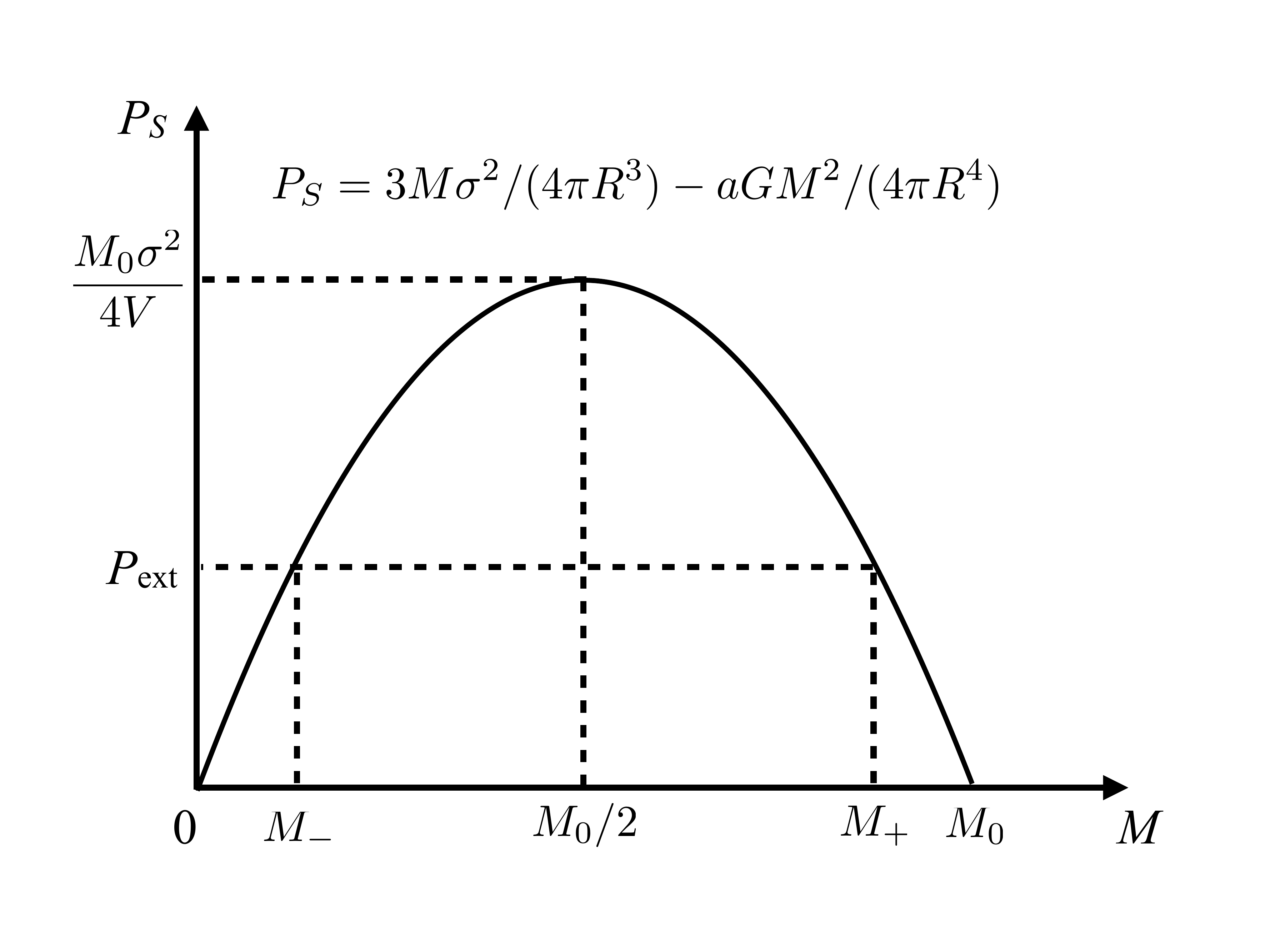}
\caption{Surface pressure $P_S$ as a function of mass $M$ for
a spherical, virialized cloud with constant radius $R_c$ and
internal velocity dispersion $\sigma$. Given an arbitrary
external pressure $P_{\rm ext}$ between $0$ and the characteristic pressure
$M_0\sigma^2/(4V)$,
two solutions exist for $M$ that allow the cloud to be in hydrostatic equilibrium,
i.e. $P_S=P_{\rm exp}$. One solution, $M_+$, which is unstable (see text), falls
in the mass range between the commonly used virial mass, $M_0=3R\sigma^2/(aG)$,
and half of it, $M_0/2$. The other solution, $M_-$, is stable and has a maximum
value of $M_0/2$.
\label{fig1}}
\end{figure}

The connection between the above Virial theorem and the external pressure
$P_{\rm ext}$ comes through the surface pressure $P_S$ -- a cloud in hydrostatic
equilibrium with the surrounding medium will have $P_S=P_{\rm ext}$.  When
$P_{\rm ext}=0$, Equation~(\ref{eq:virial1}) has only one non-trivial solution
$M_0 := M =3R\sigma^2/(aG)$, that is {\em the commonly-used virial mass} in
the literature \citep[e.g.,][]{Zwicky37}. It can be easily verified that $M_0$ is essentially the same as
the Jeans mass $M_J=\lambda_J^3\rho$ (differ by only a factor of $1.02$), where
$\lambda_J=\sigma\sqrt{\pi/(G\rho)}$ is the Jeans length.

In a more general situation, which is what defines the fundamentals of this research, $P_{\rm ext}>0$, and then two non-trivial
solutions exist for $M$ in Equation~(\ref{eq:virial1}), as is illustrated
in Figure~\ref{fig1}.  The two solutions $M_\pm$ \citep[and we note that we adopt the
notation of][ with $M_+\ge M_-$]{oka01}, can be expressed in a compact format

\begin{align}
M_\pm&=\alpha_\pm M_0,~{\rm with}~\alpha_\pm:=\frac{1\pm\sqrt{1-\beta}}{2},\label{eq:Mpm}
\end{align}

\noindent
where we have defined the dimensionless mass $\alpha:=M/M_0$ and the dimensionless
surface pressure $\beta:=4P_S V/(M_0\sigma^2)$, where $V= 4\pi R_c^3/3$
is the volume of the cloud.

\begin{figure}
\includegraphics[width=84mm]{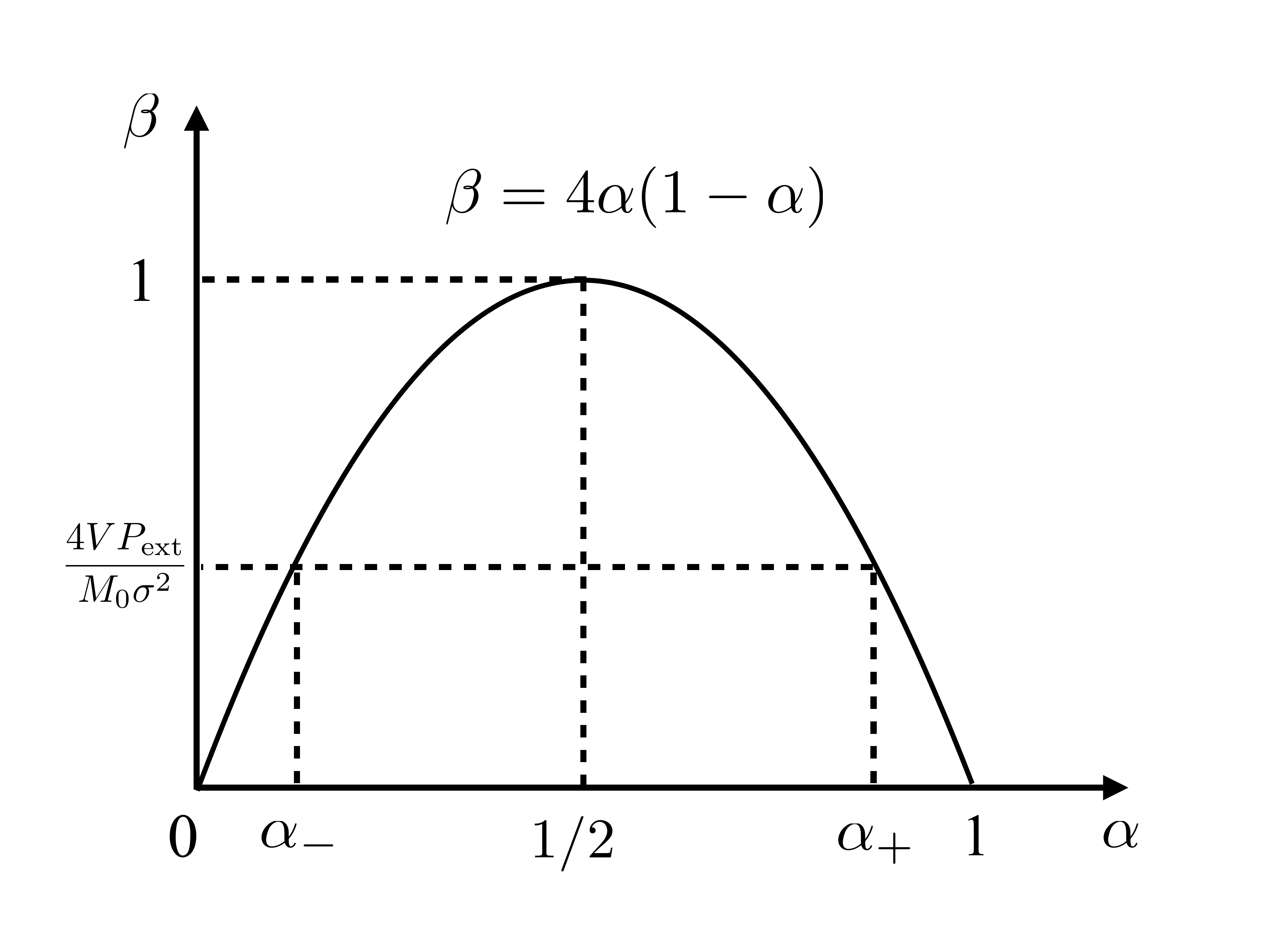}
\caption{Same as Figure~\ref{fig1} but plotting the
dimensionless quantities. The mass $M$ of a cloud is normalized by the
commonly-used virial mass $M_0$, so that $\alpha=M/M_0$. The surface pressure
$P_S$ is normalized by the characteristic pressure $M_0\sigma^2/(4V)$, so
$\beta=4P_SV/(M_0\sigma^2)$.
\label{fig2}}
\end{figure}

Thanks to these dimensionless parameters, we can rewrite Equation~(\ref{eq:virial1}) in
a dimensionless way: $\beta=4\alpha-4\alpha^2$, which is useful to understand the role of these
parameters in assessing the correct values for the mass of the cloud, as we will see later.
Figure~\ref{fig2} displays the geometric meaning of these dimensionless parameters.

A close analysis of Equation~(\ref{eq:virial1}) reveals crucial information about
the stability of gas clouds in galactic nuclei:

\begin{enumerate}

\item Given (for fixed) $R_c$ and $\sigma$ -- quantities that can be extracted
from observations -- the maximum $P_S$ that a cloud can have is
$M_0\sigma^2/(4V)$.  This is also the maximum external pressure that a cloud
can withstand in hydrostatic equilibrium.  At this critical point (because it is an
inflection point, see Figure~\ref{fig1}), $P_{\rm ext}=M_0\sigma^2/(4V)$, the
only solution to maintain hydrostatic equilibrium of the cloud is $M=M_0/2$.

\item When $0<P_{\rm ext}<M_0\sigma^2/(4V)$, we have that $M_0/2<M_+<M_0$ and
$0<M_-<M_0/2$.  Hence, in this regime of $P_{\rm ext}$, $M_0$ is {\em always}
an overestimation of the {\em real} mass of the cloud.

\item Only one branch of the solution for the mass of the cloud is {\em
dynamically stable}.  To see this, let us consider a cloud already in hydrostatic
equilibrium, and then we perturb it by either increasing or decreasing its mass.
(i) First branch: $M=M_+$.  If we slightly increase $M$, the surface gravity of
the cloud increases, meanwhile the surface pressure of the cloud decreases
because $dP_S/dM<0$, as shown in Figure~\ref{fig1}.  The tendency of the cloud
is to shrink, but this will increase the surface gravity even more, so the
cloud collapses. On the other hand, if we slightly decrease $M$, then the
surface gravity decreases while $P_S$ increases, so the cloud will expand,
which leads to an even smaller surface gravity, therefore the cloud eventually
explodes. (ii) However, for the second branch of the solution, $M=M_-$, we have
$dP_S/dM>0$, so the cloud is in the opposite situation -- $P_S$ changes in
proportion with the change of the surface gravity.  Now the cloud is
dynamically stable.

\item As a consequence of the last point, we can see that only $M_-$ is
physically correct. Observationally, gas clouds are often found in places where
the observed pressure (i.e. not derived from the Virial theorem) is such that
$P_{\rm ext} \ll M_0\sigma^2/(4V)$, or $\beta\ll1$ in the dimensionless
expression, as we will see later in the practical examples.  Hence, a common
practice in the literature is to use $P_S = 0$ as an approximation (without any
justification a priori). As a result, the only non-trivial solution of
Equation~(\ref{eq:virial1}) is $M= M_0$, which is the commonly-used virial
mass. Nonetheless, we know that there are, strictly speaking, two solutions for
$M$ even when $P_S$ is very small, and that $M_0$ is an approximation to the
$M_+$ solution. This $M_0$ {\em cannot} be the real mass of the cloud because
(i) it is, at best, its upper limit, (ii) it is on the {\em unstable} branch
for the solution of $M$, and (iii) connected to the first reason, it can
overestimate the cloud mass (mostly likely $M_-$) by orders of magnitude, since
$M_0/M_-=1/\alpha_-\gg1$ when $\beta\ll1$.

\end{enumerate}

Point (4) highlights the necessity of including external forces in the analysis
of the stability of clouds, no matter how small the external forces may be.
This point is often overlooked in the literature.  The root of this ignorance
is the intuition that self-gravity is the dominant binding force in clouds.

This intuition is fallacious, as one can see by evaluating the relative
importance of the two binding forces in the Virial theorem, i.e. comparing the
external pressure term $4\pi R_c^3P_S$ and the self-gravity term $aGM^2/R_c$ in
Equation~(\ref{eq:virial1}).  Their ratio, $(1-\alpha)/\alpha$, is greater than
$1$ for a dynamically stable cloud, i.e. external pressure predominates,
because we have proven that $\alpha$ must be smaller than $1/2$ to qualify as a
stable solution. If the ratio is smaller than $1$, i.e. if the self-gravity
predominates, the value of $\alpha$ cannot be elsewhere but greater than $1/2$,
and we have proven that such a solution is dynamically unstable.

\subsection{Tidal Terms}\label{sec:tidal}

A galactic nucleus, which is the focus of this work, usually has a SMBH sitting
at the centre with a dense (nuclear) star cluster surrounding it. Their
presence introduces in our problem an additional ingredient of tidal fields,
which will result in new terms to be added to the Virial theorem.

Besides feeling tidal forces, our test cloud is very likely to have a certain
degree of rotation as well. This is so because its orbital motion in the
galactic nucleus leads to periodic tidal perturbations, which in the long run
tends to synchronize the spin of the cloud with its orbital period
\citep{gladman96}. At this point, we say that the cloud is ``tidally locked''.
In this subsection, we will include the effects of rotation in our analysis by
exploring two representative cases: (i) When the cloud is non-rotating with
respect to the rest frame -- the observer on the Earth -- and (ii) when it is
already tidally locked.

In the following, we will show that (a) the Virial theorem without tidal
forces, (b) the Virial theorem with tidal forces but no rotation, and (c) the
Virial theorem with tidal forces and a tidally-locked cloud can all be merged
into a new, universal theorem.

We start our analysis by adding in Equation~(\ref{eq:eom}) an external
gravitational potential $\Phi_{\rm ext}$, accounting for both the SMBH and the
nuclear star cluster. The equation of motion of a fluid element inside our test
cloud hence becomes

\begin{align}
\rho\ddot{\vec{R}}&=-\vec{\nabla}P-\rho\vec{\nabla}\Phi_c
-\rho\vec{\nabla}\Phi_{\rm ext}.
\label{eq:eomR}
\end{align}

\noindent
In the last equation we have introduced $\vec{R}$ as a position vector for the
fluid element with its origin at the SMBH, which we assume to be at the
galactic centre. We now introduce a new position vector, $\vec{D}$, which has
the same origin but points at the Centre-of-Mass (CoM) of the cloud. The
difference of the two, $\vec{R} - \vec{D}$, is a vector that points from the
CoM to the fluid element of the cloud, which we name $\vec{r}$.
Figure~\ref{fig3} illustrates the definitions of the vectors.  These new
vectors allow us to re-write the last equation as follows:

\begin{align}
\rho\ddot{\vec{r}}&=-\vec{\nabla}P-\rho\vec{\nabla}\Phi_c
+\underbrace{\rho[-\vec{\nabla}\Phi_{\rm ext}(\vec{R})-\ddot{\vec{D}}]}_{\rm tidal~term,\,\rho\,\vec{a}_{t}}. \label{eq:eomtide}
\end{align}

\noindent
In the brackets we can now easily identify -- thanks to the introduction of
$\vec{D}$ and $\vec{R}$ -- the tidal acceleration $\vec{a}_{t}$.

We now proceed the same way as we did in Section~\ref{sec:P} -- take the dot
product with respect to $\vec{r}$ and integrate Equation~(\ref{eq:eomtide})
over the volume of the cloud. The integration is different depending on whether
the cloud is rotating or not.

\subsubsection{Non-rotating Clouds}

In this case, the orientation of $\vec{r}$ is fixed with respect to an observer
standing on the earth, so that the integration is almost straightforward.  As a
matter of fact, the integration has already been done in Section~\ref{sec:P}
(see Equations~\ref{eq:intlhs} to \ref{eq:virial1}) except for the last tidal
term $\rho\,\vec{a}_{t}$.

\begin{figure}
\includegraphics[width=84mm]{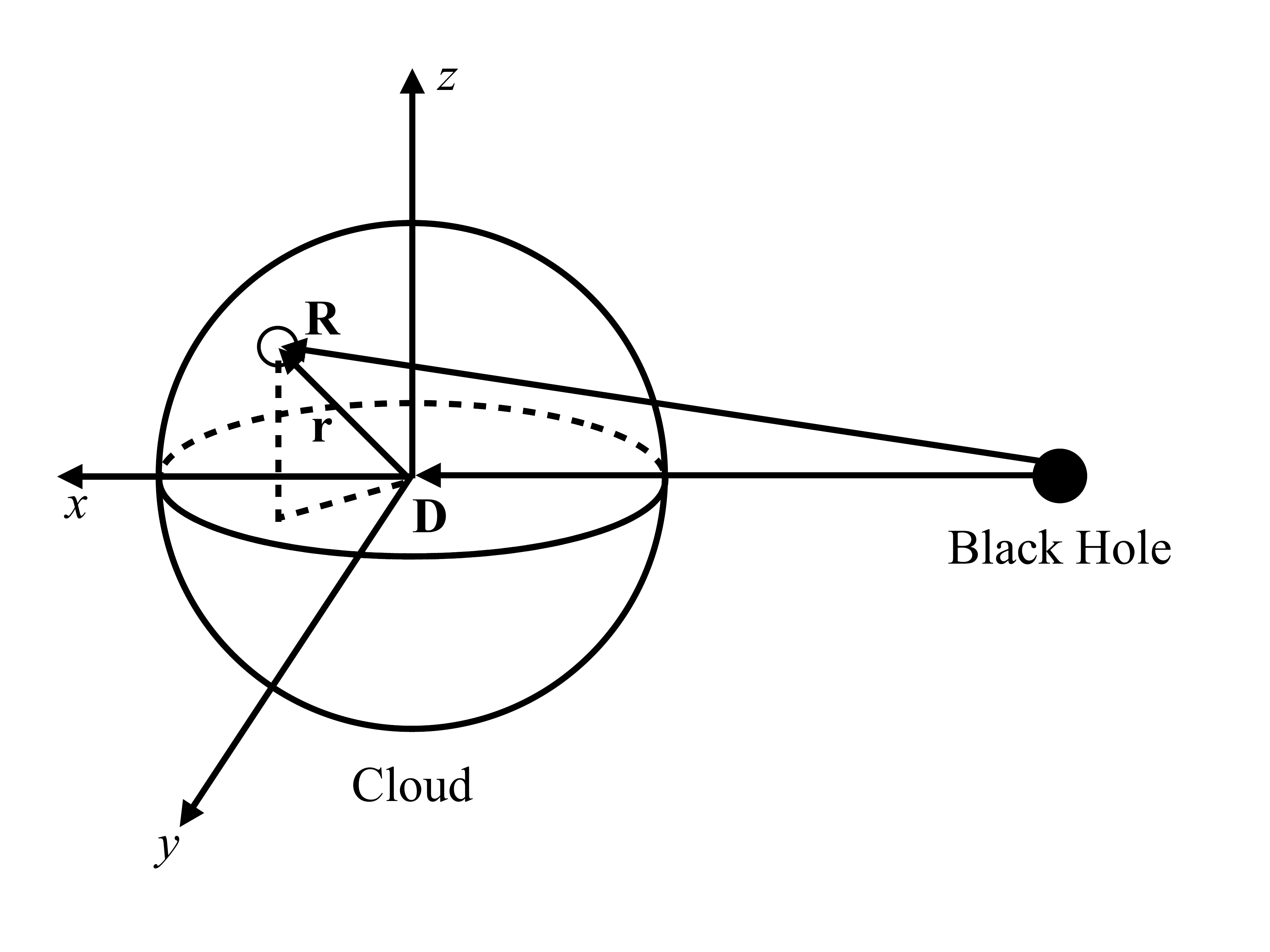}
\caption{Reference frame for a fluid element inside a non-rotating cloud which
is orbiting a SMBH. Starting from the SMBH, the vector $\vec{R}$ points to the
fluid element and $\vec{D}$ points to the CoM of the cloud.  The
coordinates are chosen in a way such that the origin coincides with the CoM
of the cloud, the $x$-axis points in the same direction as $\vec{D}$, and the
$z$-axis is perpendicular to the orbital plane of the cloud.  In this
coordinate system, the vector $\vec{R}-\vec{D}$ is pointing from the CoM of the cloud to
the fluid element, which we denote as $\vec{r}$, and its three components
are $(x,\,y,\,z)$.
\label{fig3}}
\end{figure}

To complete the integration, we need to know the three components of the vector
$\vec{a}_{t}$, so we have to define our working coordinates. As is illustrated
in Figure~\ref{fig3}, we choose the origin of coordinates at the CoM of the
cloud. The $x$-axis is chosen to be aligned with $\vec{D}$ and the $z$-axis
aligned with the {\em orbital} angular momentum of the cloud.  Hence,
$\vec{r}=(x,\,y,\,z)$.  We define $\omega_c$ to be the angular velocity of a
circular orbit at a distance of $D$ from the SMBH, so that
$\ddot{\vec{D}}=(-\omega_c^2D,\,0,\,0)$ if the nuclear star cluster is
spherically symmetric.

Given that $|\vec{r}|\ll|\vec{D}|$ in the system of our interest, we do a
linear expansion for the components of $\vec{a}_t$ and find

\begin{align}
\vec{a}_t=(xT,\,-y\omega_c^2,\,-z\omega_c^2).
\end{align}

\noindent
Here we have introduced $T=-Dd\omega_c^2/dD$, the tidal acceleration per unit
length in the radial direction \citep[e.g.][]{stark78}. The two components with
negative signs correspond to compressive tides: A fluid element moving away
from the CoM of the cloud will experience a restoring acceleration in the $y$
and $z$ directions.  Given that $T$ is positive in galaxies (as we will see
later), the acceleration in the $x$ direction is pointing away from the CoM.

Therefore, we have seen that the tidal forces acting on the cloud is not
spherically symmetric. Consequently, a cloud in hydrostatic equilibrium is not
strictly spherical.  However, for mathematical simplicity, we inherit
sphericity for the later analysis and we deem this approximation valid to
first-order.  This small sacrifice of mathematical accuracy will reward us with
useful physical insights, as we will see. Now using the three components of
$\vec{a}_t$ and assuming sphericity, we derive

\begin{align}
\int_V\vec{r}\cdot\vec{a}_tdm&=bR^2M(T-2\omega_c^2),\label{eq:introt1}
\end{align}

\noindent
where $b = 1/5$ for spherical, homogeneous clouds.

\subsubsection{Rotating, Tidally-locked Clouds}

\begin{figure}
\includegraphics[width=84mm]{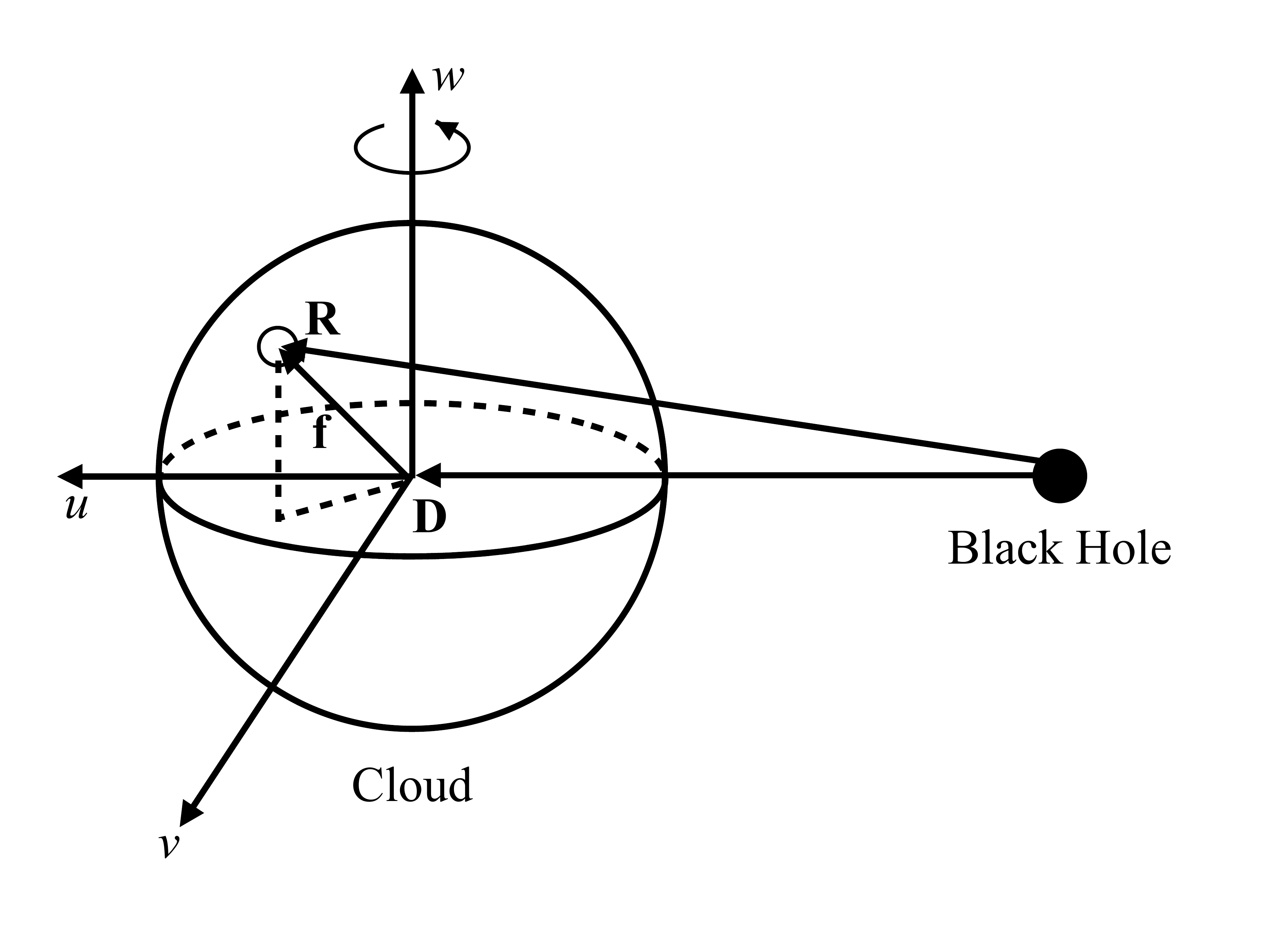}
\caption{Reference frame for a fluid element in a rotating cloud. The two
position vectors, $\vec{R}$ and $\vec{D}$, and the origin of the coordinates
are defined in the same way as in Figure~\ref{fig3}, but the orientations of the
coordinates are not. Here, the $w$-axis is defined by the rotating axis of the
cloud, and the other two orthogonal axes, namely $u$- and $v$-axes, are
co-rotating with the cloud. Now we denote the vector $\vec{R}-\vec{D}$ as
$\vec{f}$, to distinguish it from the $\vec{r}$ vector in the non-rotating
case, and in the current coordinate system $\vec{f}$ has three components
$(u,\,v,\,w)$. If the cloud is tidally locked, the
$w$-axis will be perpendicular to the orbital plane of the cloud.
\label{fig4}}
\end{figure}

The second representative case in our study is a cloud that is tidally locked
in its orbit around the SMBH.  In this case, the integral of the LHS of
Equation~(\ref{eq:eomtide}), i.e. Equation~(\ref{eq:intlhs}), is no longer
zero. To address the calculation, we now choose a more convenient frame, the
one centered on the CoM of the cloud and co-rotating with it. Figure~\ref{fig4}
illustrates this co-rotating frame.

From the CoM to a fluid element in the cloud, the position vector is defined to
be $\vec{f}:=u\,\vec{e}_u+v\,\vec{e}_v+w\,\vec{e}_w$, with $\vec{e}_i$
($i\equiv u,\,v,\,w$) the unity vector of components.  As in the previous
non-rotating case, the relation $\vec{R}=\vec{D}+\vec{f}$ still holds.
Therefore, starting from Equation~(\ref{eq:eomR}), we derive the equation of
motion for the considered fluid element as:

\begin{align}
\rho\,\ddot{\vec{f}}&=-\vec{\nabla}P-\rho\vec{\nabla}\Phi_c
+\rho\,\vec{a}_t. \label{eq:eomrot}
\end{align}

This equation is similar to Equation~(\ref{eq:eomtide}), except that the unit
vectors that we choose for $\vec{f}$ are rotating respect to an observer on
Earth.  As a result of this choice, $\ddot{\vec{f}}$ is a combination of two
accelerations, i.e.

\begin{align}
\ddot{\vec{f}}=\underbrace{(\ddot{u},\,\ddot{v},\,\ddot{w})}_{\rm In~the~rotating~frame}-\underbrace{\vec{a}_{\rm cen}}_{\rm centrifugal},
\end{align}

\noindent
where the acceleration $\vec{a}_{\rm cen}$ comes from the usual concept of a
centrifugal force. Let us consider now the consequences for the integration:

\begin{itemize}

\item
The first acceleration corresponds to the motion of a fluid element with
respect to the chosen frame. Because we have chosen the co-rotating frame, the
fluid elements of the cloud are ``frozen'' with respect to the frame, so an
integration of the dot product of this acceleration with $\vec{f}$ will vanish,
exactly as in Equation~(\ref{eq:intlhs}).

\item
As for the centrifugal acceleration, the integral
$\int_V\vec{f}\cdot\vec{a}_{\rm cen}dm$ does not vanish because the product
$\vec{f}\cdot\vec{a}_{\rm cen}$ is always positive, except for those fluids
lying on the rotation axis of the cloud, in which case
$\vec{f}\cdot\vec{a}_{\rm cen}$ will be zero.

\end{itemize}

After these considerations, we can rearrange Equation~(\ref{eq:eomrot}), moving
all the terms with non-vanishing integrals to the RHS, and derive a new, more
convenient form for the equation of motion:

\begin{align}
\rho\,(\ddot{u},\ddot{v},\ddot{w})&=-\vec{\nabla}P-\rho\vec{\nabla}\Phi_c
+\rho\,(\vec{a}_t+\vec{a}_{\rm cen}). \label{eq:eomrot2}
\end{align}

\noindent
We are now left with only one task, namely the integration of the tidal term
$\int_V\vec{f}\cdot(\vec{a}_t+\vec{a}_{\rm cen})\,dm$, because we already know
the outcome of the other integrals from the previous subsection.

If the cloud is tidally-locked and on a circular orbit around the SMBH, the
angular velocity (or spin) vector  $\vec{\Omega}=(0,\,0,\,\omega_c)$ will be
constant, and the centrifugal acceleration can be calculated with $\vec{a}_{\rm
cen}=-\vec{\Omega}\times(\vec{\Omega}\times\vec{f})=(u\omega_c^2,\,v\omega_c^2,\,0)$.
After some algebra we find

\begin{align}
\int_V&\vec{f}\cdot(\vec{a}_t+\vec{a}_{\rm cen})dm=bR^2MT,\label{eq:introt2}
\end{align}
where, again, we assume sphericity for the cloud.

\subsection{The Extended Virial Theorem}\label{sec:eVirial}

Before we continue, we recapitulate what we have got so far: (i) The
commonly-used virial mass $M_0$ is inevitably an overestimation of the mass of
a dynamically stable cloud. (ii) The tidal forces lead to an extra term in the
derivation of the Virial theorem, and we shall call this new theorem the {\em
extended Virial theorem} (EVT hereafter). (iii) We evaluated two representative
cases to calculate this extra term, namely a non-rotating cloud, which led to
Equation~(\ref{eq:introt1}), and a tidally-locked cloud, which resulted in
Equation~(\ref{eq:introt2}).

It is easy to see that the EVT takes the form:

\begin{align}
4\pi R_c^3P_S&=3M\sigma'^2-aGM^2/R_c,\label{eq:virial2}
\end{align}

\noindent
if we define $\sigma'$, the effective velocity dispersion, in the following way:

\begin{align}
\sigma'^2&=\sigma^2+bR^2(T-2\omega_c^2)/3
\end{align}

\noindent
for non-rotating clouds and

\begin{align}
\sigma'^2&=\sigma^2+bR^2T/3
\end{align}

\noindent
for tidally-locked ones. Now it is clear that an external tidal field
effectively changes the internal energy of a cloud, in other words, it changes
the ability of the cloud to expand or to contract.

The discussion presented in Section~\ref{sec:P} must be re-evaluated in the
light of our newly derived Virial theorem. Here we highlight two crucial
points.  (i) It can readily be seen that Equation~(\ref{eq:virial2}) is
structurally identical to Equation~(\ref{eq:virial1}). The only difference is
the newly defined $\sigma'$, which equivalently introduces a new virial mass,
$M'_0=3R\sigma'^2/(aG)$. Correspondingly, the dimensionless mass should be
redefined as $\alpha'=M/M'_0$ and the dimensionless surface pressure as $\beta'
= 4P_S V/(M'_0\sigma'^2)$, so that the EVT retains a dimensionless form of
$\beta'=4\alpha'(1-\alpha')$.  (ii) The ETV is a quadratic function of $M$,
same as the conventional Virial theorem.  Therefore, we can prove that given
$\beta'$, the dynamically-stable solution is $M'_-=\alpha'_-M'_0$, where
$\alpha'_-<(1/2)$.  Again, this solution stands for a cloud bounded mainly by
the external forces.

We now rewrite $\sigma'$ in a different way, so that it is easier to understand
what the implications of our EVT are. If the central SMBH
has a mass $M_\bullet$ and the nuclear star cluster follows a density profile
$\rho_*\propto D^{-\gamma}$ \citep[see][ for a review]{ama12}, the
angular velocity $\omega_c$ of a circular orbit can be written as

\begin{align}
\omega_c^2=\frac{G[M_\bullet+M_*(D)]}{D^3},
\end{align}

\noindent
where $M_*(D)\propto D^{3-\gamma}$ is the stellar mass enclosed by the orbit.
It follows that

\begin{align}
\sigma'^2& = \sigma^2+
\frac{bR^2}{3}\left[\underbrace{\frac{GM_\bullet}{D^3}}_{\rm black~hole}
-\underbrace{(2-\gamma)\frac{GM_*(D)}{D^3}}_{\rm stellar~component}\right]\label{eq:sigprim1}
\end{align}

\noindent
for the non-rotating clouds

\begin{align}
\sigma'^2& = \sigma^2+
\frac{bR^2}{3}\left[\frac{3GM_\bullet}{D^3}
+\gamma\frac{GM_*(D)}{D^3}\right]\label{eq:sigprim2}
\end{align}

\noindent
for tidally-locked clouds. It is clear that, as marked in
Equation~(\ref{eq:sigprim1}), two components contribute to the tidal effects.
The first contribution is directly linked to the mass of the SMBH, and the
second one to the mass {\em and shape} of the stellar system.

A close inspection of the last two equations reveals two pieces of new information:

(i) If the cloud is non-rotating, then we could have $\sigma'<\sigma$, i.e.
when the SMBH term of Equation~(\ref{eq:sigprim1}) becomes smaller than the
stellar one.  In this case the tidal force is effectively reducing the internal
energy of the cloud. This effect is coming from the stellar component,
especially its tidal forces in the tangential direction which are acting in a
compressive way on to the cloud.  On the other hand, we can have the inverse
situation, in which $\sigma<\sigma'$, i.e. when the cloud is so close to the
SMBH so that the role of the stellar system can be neglected (very small $D$).
In this latter situation, the tides, induced mainly by the SMBH, essentially
are tearing the cloud apart, increasing the internal energy of the cloud.

(ii) In the tidally-locked case, we have a similar situation, because $\gamma$
is not necessarily always positive \citep[e.g.][]{merritt10}. Contrasting the
last two equations, one can see that a cloud that is tidally-locked always has
a larger $\sigma'$ than that of a non-rotating one. This excess of effective
internal energy comes from the additional centrifugal acceleration.

\section{Applications}\label{sec:app}

In this section we apply the stable solution from the EVT, i.e.
$M'_-=\alpha'_-M'_0$, to infer the physical properties of observed gas clumps
in two types of galactic nuclei, namely AGNs and our GC.  As we
will see, using $M'_-$ instead of the conventional virial mass $M_0$ resolves a
series of controversies and provides interesting implications for the origin
and evolution of these clumps.

From now on we will call an observed object a ``clump'', to distinguish it from
the idealized, theoretical concept of a ``cloud''. This distinction is meaningful
because it draws attention to the limitations of the observational data -- while
clouds have well-defined shapes and physical parameters, clumps do not have
clearcut boundaries, nor are all their physical quantities observable.

In each of the three subsections that follow we (i) first introduce the current
controversy linked to the usage of the conventional virial mass and (ii)
resolve the controversy by applying our EVT.

\subsection{Clumps in AGN Tori}\label{sec:tori}

{\em Controversy: } Analysis of the X-ray variability of AGNs indicates the
existence of a population of dusty clumps that are very close to the central
SMBHs \citep[see][ for a review]{netzer15}. The location of these clumps
coincides approximately with the inner boundary of the dusty torus
\citep[e.g.][]{elvis00,risaliti02}.  Recent work, such as the one by
\citet{markowitz14}, provides us with the required observable quantities for a
clump.  We summarize here the typical values for them: (i) A size of
$R_c\simeq2.0\times10^{14}~{\rm cm}$, (ii) a Hydrogen (number) density of about
$n_{\rm H}\simeq3.5\times10^{8}~{\rm cm^{-3}}$, (iii) a mass for the central
SMBH of $M_\bullet\simeq4.2\times10^7~M_\odot$, (iv) a distance of $D\simeq
5.0\times10^{17}~{\rm cm}$ from the SMBH , and (v) a bolometric luminosity of
$L\simeq1.6\times10^{44}~{\rm erg~s^{-1}}$ for the AGN.  From the size and
density, we derive a typical mass of $M=10^{-5}~M_\odot$ for a spherical and
homogeneous clump.

Naively, one would expect a clump to be tidally dissolved if the tidal force
from the SMBH acting on to the clump exceeds its self-gravity. Hence, it would
seem that the density of a clumps should be at least

\begin{align}
n_{\rm tide}&\simeq M_\bullet/(m_pD^3).
\end{align}

\noindent
From the quoted typical values and a proton mass of $m_p=1.67\times10^{-24}$ g,
this critical density is $n_{\rm tide} \simeq 4.0\times10^{11}~{\rm cm^{-3}}$,
a thousand times above $n_{\rm H}$. And yet the clumps exist. This paradox led
\citet{markowitz14} to discuss possible stabilizing mechanisms, including
magnetic field and external pressure.

{\em Solution: } The aforementioned clump -- whose self-gravity is weaker than
the external tidal forces but remains bound and stable -- is exactly the type
of cloud we have found as the stable solution of our EVT
(Section~\ref{sec:eVirial}).  Our EVT predicts that the clump has a
non-negligible surface pressure $P_S$, and its mass is $\alpha'_-M'_0$ where
$\alpha'_-<(1/2)$. Therefore, we equate $\alpha'_-M'_0$ to $10^{-5}~M_\odot$,
the typical mass of the clump, and we proceed to calculate $\alpha'_-$ and then
$P_S$.

To do so, we have all the data we need except those of $\sigma$, $\gamma$, and
$M_*(D)$.  However, it is relatively straightforward to obtain a
well-established value for $\sigma$.  \cite{krolik89} proved that the typical
gas temperature in a molecular cloud in an AGN torus is about $10^3$ K.  This
corresponds to a typical internal velocity of $\sigma \approx 3~{\rm
km~s^{-1}}$.

As for $\gamma$ and $M_*(D)$, which determine the value of $\sigma'$, we notice
that $M_*(D)\ll M_\bullet$ at the distance $D\simeq 0.2$ pc where the clumps
are observed.  Therefore, we can neglect the stellar components, i.e.  those
determined by  $M_*(D)$ and $\gamma$, in the analysis of the tidal terms.

Although the SMBH term predominates in the brackets of
Equations~(\ref{eq:sigprim1}) and (\ref{eq:sigprim2}), the two terms in the
brackets are multiplied by $R_c^2$.  This multiplication leads to a negligibly
small contribution compared to $\sigma^2$, so essentially $\sigma'
\approx\sigma\approx 3~{\rm km~s^{-1}}$.  It is important to know in advance
that (i) the tidal term from the SMBH is not always negligible, as we will see
in Section~\ref{sec:CND}, and (ii) neither is the stellar component always
negligible, as is shown in Section~\ref{sec:CMZ}.

We have specified the values of all the terms entering the EVT, so we can solve
$\alpha'_-$ and $P_S$. From $\sigma'$ and $R_c$ we derive
$M_0'\simeq0.67~M_\odot$. It follows that
$\alpha'_-=M/M'_0\simeq1.5\times10^{-5}$.  Since the value we have just derived
for $\alpha'_-$ is smaller than $1/2$, the mass $M=10^{-5}~M_\odot$ qualifies
as a stable solution in the context of Equation~(\ref{eq:virial2}).  Since we
know that $\alpha'_-$ must satisfy the relationship
$\beta'=4\alpha'_-(1-\alpha'_-)$, we derive that
$\beta'\simeq4\alpha'_-\simeq6.0\times10^{-5}$, and from the definition of
$\beta'$ (Section~\ref{sec:eVirial}), we finally find
$P_S\simeq5.4\times10^{-5}~{\rm erg~cm^{-3}}$.

Now that we have the value for the surface pressure of a stable {\em cloud} in
our theory, we calculate with observed data the typical external pressure of
the environment where real clumps are found and compare the two.

The main source of external pressure is the {\em turbulent} gas surrounding the
clump. Estimating its magnitude requires knowledge of the mean density $n_{\rm
ext}$ and the turbulent velocity $v_{\rm ext}$ of such external gas medium. To
estimate $v_{\rm ext}$, we recall the following two facts. (i) AGN tori are
dusty \citep{netzer15} and dust grains will be destroyed if the collisional
velocity induced by turbulence exceeds $120~{\rm km~s^{-1}}$ \citep{shull78}.
(ii) The turbulent velocity must also be comparable to the Keplerian velocity,
about $10^2$ to $10^3~{\rm km~s^{-1}}$ depending on $M_\bullet$ and $D$, to
maintain the empirically required geometric thickness of the torus
\citep{krolik88}.  Taking both facts into account, we adopt $v_{\rm
ext}=100~{\rm km~s^{-1}}$.

For $n_{\rm ext}$, we know that  the integrated Hydrogen column density in the
mid-plane of the torus is an observable, and the typical value is $N_H \sim
10^{24}\,{\rm cm}^{-2}$ \citep[][]{buchner15}.  If we estimate the width
$\Delta D$ of the torus -- measured from the inner boundary to the outer one of
the torus -- we can derive $n_{\rm ext}$ approximately by $N_H/(\Delta D)$. To
estimate $\Delta D$, we use two pieces of empirical information: (i) there is a
correlation between AGN luminosity and the radius of the inner boundary of the
torus \citep{nenkova08}, according to which the inner boundary is at $0.2$ pc
for our typical AGN with a luminosity of $L\simeq1.6\times10^{44}~{\rm
erg~s^{-1}}$, and (ii) the outer boundary is typically  $10$ times more distant
than the inner one \citep{elitzur07}, so it is at $2$ pc in our case.  As a
result, we find $\Delta D\simeq1.8$ pc, and hence $n_{\rm
ext}\simeq1.8\times10^{5}~{\rm cm^{-3}}$.

With the numbers that we have derived for $v_{\rm ext}$ and $n_{\rm ext}$, we
proceed to calculate the external pressure  with $P_{\rm ext}=m_p \,n_{\rm
ext}\,v^2_{\rm ext}$, and the result is $2.9\times10^{-5}~{\rm erg~cm^{-3}}$.
We find that $P_{\rm ext}$ is consistent with $P_S$ within a factor two.  This
agreement supports our proposal that the clumps are confined by an external
pressure and are dynamically stable.

We note that the radiation from the central AGN cannot be another source of
external pressure. This is so, because a dusty cloud exposed to AGN irradiation
will evaporate within one orbital period \citep{pier95,namekata14}. The
radiative pressure would be

\begin{align}
P_{\rm rad}=L/(4\pi D^2c),
\label{eq:Prad}
\end{align}

\noindent
with $c$ the speed of light. We can see that this value is
$5.1\times10^{-3}~{\rm erg~cm^{-3}}$, more than two orders of magnitude higher
than the surface pressure $P_S$ derived above for the clump. Hence, for the
clump to not be dissolved, it must be shielded during most of its orbital phase
by the gas or other clumps located closer to the central SMBH, as has been
suggested by \citet{krolik88,namekata14}. Our result, hence, supports this
shielding.

\subsection{Clumps in the CND of the Milky Way}\label{sec:CND}

{\em Controversy: }
Molecular clumps have been detected in the centers of quiescent galaxies as
well \citep{hunt04,mazzalay13,davies14}.  In the GC, in particular, tens of
molecular clumps reside in the CND, a ring-like (narrow projected width)
structure surrounding Sgr A*. The observable parameters and their typical
values are $D=1.8$ pc, $R_c=0.25$ pc, and $\sigma=11~{\rm km~s^{-1}}$
\citep[from][]{christopher05}, and the central SMBH has a well-established mass
of $M_\bullet \approx 4\times10^6~M_\odot$ \citep{genzel10}.

There exists two different methods to derive the masses of these clumps.
However, both lead to dilemmas that seem to invalidate the method. We now
explain the methods and the problems that emerge with them.

(i) The typical mass of the clump according to the conventional Virial theorem,
$M_0=3R\sigma^2/(aG)$, is about $3.4\times10^4~M_\odot$ \citep[also
see][]{shukla04,christopher05,mc09,tsuboi12}. A clump with this mass has a
density of $2.3\times10^{7}~{\rm cm^{-3}}$, comparable to the tidal density
$n_{\rm tide}\simeq2.8\times10^{7}~{\rm cm^{-3}}$ at the observed distance
($D$), so the clump can withstand the tidal force. This argument has been used
to support the validity of taking $M_0$ as the real clump mass.  However, this
mass has two problems. (a) In the context of the conventional Virial
theorem, such clumps are gravitationally unstable, as discussed in points 3 and
4 of the list in Section~\ref{sec:P}. They will disappear on a free-fall
timescale (about $10^4$ yr), shorter than the required time to form an
axisymmetric CND (about $10^5$ yr, comparable to the orbital period).  In the
EVT, however, tidally-locked clumps with this mass are stable, as we will see
later.  (b) Nonetheless, $M_0$ is likely to be ruled out, because if this were
indeed the real mass of the clumps, then the CND would have a total mass of
about $10^6~M_\odot$. Such a heavy disk would inevitably destroy the {\em
observed} young stellar disk within $0.5$ pc around Sgr A* \citep[][]{subr09}.

(ii) Another way of deriving the mass of clumps is based on the assumption of
local thermal equilibrium (LTE, and in this case we call the mass derived
$M_{\rm LTE}$).  Several authors obtained a similar result, typically $M_{\rm
LTE}\simeq 2.6\times10^{3}~M_\odot$, {\em one order of magnitude smaller than
before} (from \citealt{tsuboi12}, also see
\citealt{marr93,marshall95,christopher05,liu13,rt12}). In this case, $M_{\rm
LTE}$ does not lead to the aforementioned problem related to the existence of
the young stellar disk in our GC. However, according to the naive tidal limit
criterion, these clumps will disappear within a relatively short time, in any
case shorter than the required time to form an axisymmetric CND, as has been
noticed by these authors.

{\em Solution: }
We will see that in the context of the EVT, $M_0$, as derived in (i), is
unstable for non-rotating clumps but stable for tidally-locked ones.
Nevertheless this does not change the fact that this mass leads to a scenario
ruled out by observations. On the other hand, a clump with mass $M_{\rm LTE}$,
as derived in (ii) can be stable. Since this mass does not lead to
contradictions with observations of the GC, it is a realistic value for clumps
in the CND.

We now derive the value of $P_S$ as provided by the EVT, and then we will
compare it with the environmental pressures independently derived in other
means, to prove that the clump with a mass of $M_{\rm LTE}$ is in hydrostatic
equilibrium.

The GC nuclear cluster has $\gamma=1.75$, and the enclosed stellar mass is approximately

\begin{align}
M_*(D)\simeq1.6\times10^{6}D_1^{1.25}.\label{eq:Mstar}
\end{align}

\noindent
This stellar distribution is adopted from the early work of
\citet{vollmer00,christopher05} for the sake of comparison, but it is not
significantly different from the distribution derived from more recent
observations \citep{genzel10}.  With this stellar component, we find
$\sigma'=14~{\rm km~s^{-1}}$ for a non-rotating clump and $\sigma'=19~{\rm
km~s^{-1}}$ for a tidally-locked one.

For the non-rotating clump, we hence derive $M'_0\simeq5.9\times10^4~M_\odot$,
$\alpha'_-=M_{\rm LTE}/M'_0\simeq0.044\ll1$,
$\beta'\simeq4\alpha'_-\simeq0.18$, and finally $P_S\simeq6.7\times10^{-6}~{\rm
erg~cm^{-3}}$. For the tidally-locked one, $M'_0=1.1\times10^5~M_\odot$,
$\alpha'_-\simeq0.024$, $\beta'\simeq0.094$, and
$P_S\simeq1.2\times10^{-5}~{\rm erg~cm^{-3}}$. In the two cases, we have
$\alpha'_-<(1/2)$, so we can conclude that $M=M_{\rm LTE}$ is a stable
solution. We also note that $M=M_0$ is an unstable solution for non-rotating
clumps, but it could be a stable solution for tidally-locked ones, although, as
mentioned before, this mass is excluded because of other arguments not related
to the EVT.

Balancing this surface pressure requires an environmental pressure of at least
the same order of magnitude.  In the GC we have three possibilities: First, the
hot, ionized gas in the innermost pc of the GC has a density of $10^4~{\rm
cm^{-3}}$ \citep{jackson93} and a temperature of $7,000$ K \citep{roberts93},
which provides a thermal pressure of $\sim10^{-8}~{\rm erg~cm^{-3}}$. This
pressure is too weak to explain the magnitude of $P_S$. Any derivation of the
mass of the clumps based on this pressure will lead to a too small value, such
as the $15~M_\odot$ derived by \citep{vollmer01}.  Second, at a distance of
$D=1$ pc, the ram pressure caused by the stellar winds from the observed young
stars is about $1.1\times10^{-7}~{\rm erg~cm^{-3}}$ \citep{yusef93}, which is
still too weak.  The last possibility offers a plausible solution: The
molecular gas in the CND, i.e. the ``inter-clump'' gas,  has a turbulent
velocity of $55~{\rm km~s^{-1}}$ \citep{gusten87} and a mean number density of
$10^5~{\rm cm^{-3}}$ \citep{genzel85}. The corresponding pressure due to
turbulent motion is $5\times10^{-6}~{\rm erg~cm^{-3}}$, consistent with $P_S$
as derived above.  Hence, the clumps in the CND can be stabilized by the
inter-clump turbulent gas.

The turbulence of the inter-clump gas has been postulated to be generated by a
recent outburst of Sgr A* and continuously replenished by the dissipation of
the differential rotation of the CND \citep{gusten87}. We have now a number of
evidences suggesting that the outburst took place about $6$ Myr ago and the
bolometric luminosity was as high as $10^{43}~{\rm erg~s^{-1}}$
\citep[e.g.][]{nay05,su10,bh13,ama14,chen14,chen15}.  This luminosity has a
direct, incident radiative pressure on the inner rim of the CND (at about $D=1$
pc), of about $3\times10^{-6}~{\rm erg~cm^{-3}}$ (see Equation~\ref{eq:Prad}).
This value is comparable to $P_S$ as well as to the turbulent gas pressure.
Therefore, we infer that during the outburst of Sgr A*, the inner rim of the
CND was pressurized by the radiation and that the pressure propagated into the
CND to stabilize the clumps in the context of the EVT.

\subsection{Clumps in the CMZ}\label{sec:CMZ}

{\em Controversy: }
The methods (i) and (ii) described in Section~\ref{sec:CND} have been applied
to the clumps in the CMZ of the Milky Way to derive their masses. The result is
that the virial mass $M_0$ is on average $10$ times larger than $M_{\rm LTE}$
\citep{miyazaki00,oka01}, a difference as large as we have seen in
Section~\ref{sec:CND}.

\cite{miyazaki00} investigated the possibility that the clumps in the CMZ are
confined by the hot, inter-clump ISM. However, the pressure of the hot ISM,
obtained from the observations of the $6.7$ keV iron lines, is ten times
smaller than the surface pressure that one derives based on the conventional
Virial theorem (not the EVT). Therefore, the external pressure could not explain
the discrepancy.

{\em Partial solution: }
We adopt now $D=100$ pc, $R_c=10$ pc, and $\sigma=10~{\rm km~s^{-1}}$ as the
fiducial values \citep[from][]{oka01}, but we note that the CMZ spans a large
radial range, and the clumps in it have a broad distribution in size as well as
in internal velocity dispersion.

In Section~\ref{sec:eVirial} we addressed two particular cases, namely
tidally-locked and non-rotating clumps. In the following, we will show that it
is the case of non-rotating clumps what (partially) solves the present problem
related to the clumps in the CMZ.

It does so, because the effective velocity dispersion $\sigma'$ is smaller than
the conventional one $\sigma$. Consequently, applying the Virial theorem
without accounting for the tidal forces will lead to an overestimation of the
surface pressure of the clump. We can see this by comparing the conventional
and the extended Virial theorems, i.e. Equations~\ref{eq:virial1} and
\ref{eq:virial2}. Physically, this can be envisaged as the result of ignoring
the contribution of the compressive tides in the stabilization of the clumps.

According to Equations~(\ref{eq:sigprim1}) and given that $2-\gamma>0$ in the
Milky Way \citep{genzel10}, the situation $\sigma'<\sigma$ will occur, and it
happens when the (effective) stellar mass $(2-\gamma)M_*(D)$ enclosed by the
orbit of the clump exceeds the mass $M_\bullet$ of the SMBH. The corresponding
region is at $D\gtrsim 6.3$ pc, which encompasses the entire CMZ.

To be more quantitative, we  scale Equation~(\ref{eq:sigprim1}) using the
parameters relevant to the clumps in the CMZ \citep[also from][]{oka01} and we
find

\begin{align}
\sigma'^2\simeq\sigma^2-(4.3~{\rm km~s^{-1}})^2~b\left(\frac{R_c}{10~{\rm pc}}\right)^2
\left(\frac{D}{100~{\rm pc}}\right)^{-1.75}.\label{eq:sigCMZ}
\end{align}

\noindent
Given $\sigma\sim10~{\rm km~s^{-1}}$ and $D\sim10^2$ pc, the last equation
suggests that if a clump is bigger than $10$ pc, its surface pressure will be
significantly weaker than what was derived in the early work based on the
conventional Virial theorem.

In a survey of the CMZ \citep{oka01}, $67$ out of $165$ molecular clouds have
been observed to have a size larger than $10$ pc.  Limited to these
observations, our EVT {\em partially} resolves the controversy.  Other
explanation must be sought to fully address the problem (see
Section~\ref{sec:dis} for discussion).

\section{Discussions and Conclusion}\label{sec:dis}

In this paper we analyze the stability of the gas clouds in the central
$1-10^2$ pc of a galaxy, in the region around the SMBH and the nuclear star
cluster.  We have shown that the external forces -- external pressure
(Section~\ref{sec:P}) and tidal forces (Section~\ref{sec:tidal}) -- play a
crucial role in confining and stabilizing the clouds. Based on this
understanding, we formulated an {\em extended Virial theorem} (or EVT) to
identify the correct ranges of physical quantities that lead to stability
(Section~\ref{sec:eVirial}).  We applied our EVT to model observational data
and have solved practical problems related to the stability of those gas clumps
detected in AGN tori (Section~\ref{sec:tori}), the CND in the GC
(Section~\ref{sec:CND}), and the CMZ of the Milky Way (Section~\ref{sec:CMZ},
in this last case we partially resolved the problem).

\subsection{Validity of the Assumption of Static Equilibrium}

In the previous sections we have relied on the assumption that the clouds are
in hydrostatic equilibrium. However, there are two kinds of perturbations that
potentially could invalidate this assumption. (i) Collisions with other clouds
could drive a test cloud out of equilibrium, if the cloud number density is
high enough. We define $t_{\rm coll}$ as the collisional timescale -- the mean
timespan during consecutive collisions. (ii) The effect of inhomogeneity in the
ISM can either compress or decompress a test cloud.  The fastest relative
velocity between the cloud and the ISM is the Keplerian velocity, therefore the
shortest timescale for the cloud to return to the inhomogeneous region is the
orbital period, $t_{\rm orb}$.

After one perturbation, the cloud requires at least the sound crossing
timescale $t_{\rm sc}\simeq 2R_c/\sigma$ to relax and restore equilibrium. So
our assumption of hydrostatic equilibrium is valid if these two criteria are
simultaneously met: (i) $t_{\rm sc}< t_{\rm coll}$ and (ii) $t_{\rm sc}< t_{\rm
orb}$.  We now use these two criteria to assess the destabilizing effects in
the three different media that we have considered in this paper.

(a) For a cloud in AGN torus, since we know the typical values for $\sigma$, $R_c$,
$D$, $M_{\bullet}$, we find that $t_{\rm sc}\simeq 42$ yr, $t_{\rm
orb}\simeq6.3\times10^2$ yr. Since \cite{krolik88} proved that $t_{\rm
coll}\simeq t_{\rm orb}$, we conclude that hydrostatic equilibrium can be achieved
in AGN tori.

(b) For a cloud in the CND, we find $t_{\rm sc}\simeq4.5\times10^{4}~{\rm yr}$
and $t_{\rm orb}\simeq1.1\times10^5~{\rm yr}$ using the observational data.
From the numerical simulations of \cite{vollmer02}, we have that $t_{\rm
coll}\simeq2$ Myr.  Hence, hydrostatic equilibrium can be achieved in the CND.

(c) For a cloud in the CMZ, we derive $t_{\rm sc}\simeq2$ Myr and $t_{\rm
orb}\simeq5.9$ Myr according to the observational data. For $t_{\rm coll}$, we
notice that the clumps cover about $50\%$ of the disk area \citep{oka01}, so we
use the formula for $t_{\rm coll}$ derived in \cite{bally88}, which is a
function of the covering factor, and we find $t_{\rm coll}\simeq4$ Myr for a
cloud at $D=100$ pc.  In this region, the three timescales (i.e. $t_{\rm sc}$,
$t_{\rm coll}$, and $t_{\rm orb}$) are of the same order of magnitude. As a
result, the clouds may not have had enough time to fully restore equilibrium.
This could be the reason why in Section~\ref{sec:CMZ} we only could explain a
fraction of the clumps in the CMZ. Indeed, there is evidence that at the
inner rim of the CMZ, where both $t_{\rm coll}$ and $t_{\rm orb}$ are
relatively short, the molecular clouds may have recently experienced collisions
\citep[e.g.][]{RF06}.

\subsection{Conclusion}

We have proven that the observed gas clumps in AGNs tori and in the CND of the
Milky Way (and partially in the CMZ) are not self-gravitating, but instead,
they are stabilized by external pressure and tidal forces. The standard Jeans
instability produces only self-gravitating clouds, and therefore it cannot
account for the origin of these observed clumps.

The noticeably large internal energy relative to the self-gravity, as we found
for these clumps, is more consistent with a non-conventional formation
mechanism, such as the collisional fragmentation and agglomeration
\citep{krolik88,vollmer04}. This discovery supports the picture that
clump-clump collision drives gas inflow to the central pc to power AGNs .

In the future, application of the EVT to individual clumps (i.e. observed
clouds) potentially can provide us with a mapping of clump masses and
environmental pressure around supermassive black holes in galactic nuclei.

\acknowledgments

We thank Robert Nikutta and Johannes Buchner for many discussions on AGN
torus.  XC and JC are supported by CONICYT-Chile through Anillo (ACT1101),
Basal (PFB0609),  and FONDECYT (1141175) grants.  PAS is indebt to Sonia
P{\'e}rez for her support during the write-up of the paper.

\label{lastpage}
\end{document}